\documentclass{elsart}

\usepackage{graphicx}
\usepackage{epsfig}
\usepackage{bm}
\usepackage{amssymb}

\begin{document}

\begin{frontmatter}

\title{Complex Networks on Hyperbolic  Surfaces}

\author{T. Aste, T. Di Matteo and S. T. Hyde }
\address{Applied Mathematics, Research School of Physical Sciences, Australian National University, 0200 Canberra, Australia.}

\date{\today} 


\address{\bf To appear in: { Physica A} (2004)}

\begin{abstract}
We explore a novel method to generate and characterize complex networks by means of their embedding on hyperbolic surfaces.
Evolution through local elementary moves allows the exploration of the ensemble of networks which share common embeddings and consequently share similar hierarchical properties.
This method provides a new perspective to classify network-complexity both on local and global scale.
We demonstrate by means of several examples that there is a strong relation between the network properties and the embedding surface.
\end{abstract}


\end{frontmatter}

\maketitle
\vspace{-0.5cm}
\section{Introduction}
\vspace{-0.8cm}
In recent years it has become increasingly evident that a convenient way to study systems constituted of many interacting elements is by associating to each element  a node and to each interaction a link between nodes, giving a {\it  network} (or graph). It has been widely noted that complex interconnected structures appear in a wide variety of systems of high technological and intellectual importance.
It has been pointed out that many such networks are disordered but not completely random \cite{Amaral,Barabasi,Newman,Cohen03}.
On the contrary, they have intrinsic hierarchies and characteristic organizations which are distinguishable and are preserved during the network evolution. 
In particular, one of the principal feature of these networks is the fact that they are both \emph{clustered} and \emph{connected}.
For instance, an individual in a social network has most links within his own local circle, yet each individual in the world is only at a few steps from any other \cite{WattSt98}.
An example of a completely clustered network is a triangular lattice on a planar surface: in such a network each one of the $n$ nodes is connected with its local neighbors only and the average distance between two individuals scales as $n^{1/2}$. 
This is a `large world'.
On the other hand, after Erdos and R\'enyi \cite{ER60}, we know that random graphs are closely connected systems where the average distance scales as $\ln(n)$: a `small world'.
Intermediate structures can be constructed from the planar lattice by adding links between distant nodes making in this way \emph{short cuts}.
But such an insertion of a short-cut on the triangular lattice has an important consequence: the network can no longer be drawn on the plane without edge crossings; it is non-planar.
The embedding surface must be modified accordingly by creating a `worm hole' which connects two distant parts of the surface and through which the new link can `travel'.
Such `worm holes' create short-cut tunnels in the (2D) universe transforming it into  a small world.

\begin{figure}
\begin{center}
\resizebox{0.4\textwidth}{!}{%
\includegraphics{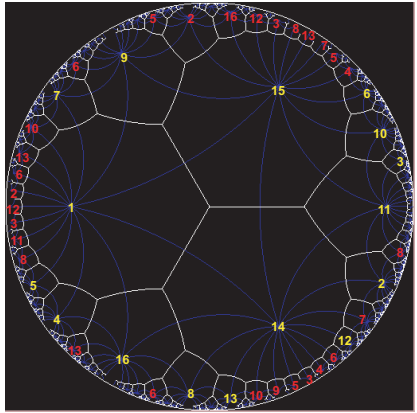}
}
\end{center}
\caption{ 
\label{f.complete}
 An embedding of $K_{16}$ (with nodes labelled from 1 to 16) formed by unfolding $S_{13}$ onto its universal cover (the hyperbolic plane) by giving multiple copies of nodes.
The dual graph   (white edges) is also drawn. 
}
\end{figure}

In this paper we explore the idea of a network that exists, grows and evolves on a hyperbolic surface.
The complexity of the network itself is in this way associated with the complexity of the surface and the evolution of the network is now constrained to a given overall topological organization.
More precisely, we explore the relation between the properties of a network and its embedding on a surface.
An orientable surface (an intersection-free, two-sided, two-dimensional submanifold of three-dimensional Euclidean space) can be topologically classified in term of its {\it genus} which is the largest number of non-intersecting simple closed cuts that can be made on the surface without disconnecting a portion (equal to the number of handles in the surface).
The genus ($g$) is a good measure of complexity for a surface: under such a classification, the sphere ($g=0$) is the simplest system; the torus is the second-simpler ($g=1$); etc.
To a given network can always be assigned a genus: defined to be equal to the minimum number of handles that must be added to the plane to {\it embed} the graph without edge-crossings. (Accordingly, a planar graph has genus 0 and it can be "minimally" embedded on the sphere).
Therefore, our approach works in two ways: it is a convenient tool to generate graphs with given complexity (genus) and/or  it is a useful instrument to measure the complexity of real-world graphs.

The aim of the present paper is to emphasize \emph{why} it is convenient to study networks in term of their embeddings on a surface.
We find several attractive features of this approach: 
1) it provides new measures to characterize complexity; 
2) it gives a locally-planar representation;
3) it provides a hierarchical ensemble classification;
4) it allows the application of topologically invariant elementary moves  \cite{p16,p17,AsteSherr}.
In addition, let us stress that \emph{any} network can be embedded on a surface, therefore:  \emph{why not?}.

\begin{figure}
\begin{center}
\resizebox{0.50\textwidth}{!}{%
\includegraphics{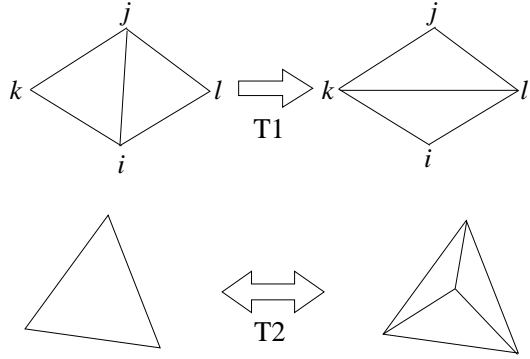}
}
\end{center}
\caption{ 
\label{f.T1T2}
Elementary moves on a triangulation: Edge switching ($T1$) and Vertex insertion and removal ($T2$).
}
\end{figure}

\vspace{-0.8cm}
\section{Hyperbolic Embeddings} \label{HG}
\vspace{-0.8cm}
In this section we discuss the embeddings of \emph{undirected}, \emph{simple}   graphs (with at most one edge  connecting any two vertices and each edge connecting two vertices) on orientable surfaces of genus $g$ .
First consider the \emph{complete} graph with $n$ vertices ($K_n$).
In $K_n$, all possible links are present and each node is connected with all other $n-1$ nodes.
Evidently, any graph with $n$ nodes is necessarily a sub-graph of $K_n$.
It is known  \cite{Ringel,p14}  that an embedding of $K_n$ is always possible on an orientable  surface $S_g$ of genus $g\ge g^*$, with
\begin{equation}
\label{g*}
g^* = \lceil{ \frac{(n-3)(n-4)}{12} }\rceil
\end{equation}
 (for $n \ge 3$, where $\lceil x \rceil$ denotes the ceiling function which returns the smallest integer number $\ge x$).
For $n > 7$ one has $g^* >1$, and hyperbolic surfaces are needed to embed the complete graph.
When $n=$ 0, 3, 4 or 7 (mod 12) the quantity $(n-3)(n-4)$ is divisible by 12 and the embedding of $K_n$ on $S_{g^*}$ is a regular  \emph{triangular} tiling.

Eq.\ref{g*} states that any graph with $n$ nodes (any sub-graph of $K_n$) can be embedded on a surface of genus larger of equal than $g^*$.
However sub-graphs of $K_n$ might be embeddable on surfaces with smaller genus.
The embedding of the complete graph on $S_{g^*}$ has the desirable feature of local planarity. Note, however, that this local simplification is achieved by introducing complexity in the global surface ($g^*$ scales with $n^2$).
Constructions of embeddings of sub-graph of $K_n$ on topologically simpler surfaces (with lower genus than $g^*$) are therefore required.

In this paper we discuss three distinct constructions: \emph{bottom-up};  \emph{top-down} and  \emph{dynamical}.
These methods (described below) are algorithmically very different, yet they can produce the same final structures.

The {\bf bottom-up} approach is  an iterative construction on a surface $S_g$ of given genus $g$. 
It starts from a set of $n$ unconnected nodes and connects two nodes if and only if the resulting graph can be embedded on $S_g$. 
This process proceeds iteratively and terminates with either a triangular graph containing $3n + 6 (g - 1)$ links (when $g < g^*$), or the complete graph $K_n$ (when $g \ge g^*$).
This construction gives \emph{maximal networks}, with a maximum number of links for a graph with $n$ vertices on $S_g$.
Less connected graphs can be constructed by edge-pruning or by ending the construction process at an earlier stage. 
Applications of this method to financial market data for the case $g=0$ are discussed in \cite{DM04interest} (interest rates data) and \cite{Michele} (stock market data).

The {\bf top-down} construction starts from the embedding of $K_n$ on $S_{g^*}$. We
then roll this multi-handled surface on its (topologically planar) \emph{universal cover}. 
Most commonly, the universal cover lies in the hyperbolic plane $H^2$ (except where $n \le 7$) \cite{p15}. 
The resulting pattern exhibits discrete symmetries, characteristic of a regular tiling of $H^2$ \cite{p15}. 
In particular the special cases $n = 0, 3, 7$ (mod 12) lead to regular tilings of $H^2$ with Schl\"afli symbols $\{3,n-1\}$ \cite{Cox} . 
This procedure effectively "unzips" the multi-handled surface, and forms multiple planar copies in the hyperbolic plane.
An example for $n=16$  (related to the case in \cite{DM04interest}) is shown in Fig.\ref{f.complete}. 
In this case $g^*= 13$ (Eq.\ref{g*})  and the figure represents the embedding of $K_{16}$ by unfolding $S_{13}$ on the universal cover. 
This is one possible embedding among the exponentially-large class of allowed embeddings \cite{Bonn00}.
Once we have formed the universal cover, we can proceed in the inverse direction by edge pruning $H^2$ followed by re-gluing the universal cover into a simpler surface, $S_{g}$
 \cite{HydeEUPJB03,HydeSSS03}.

The {\bf dynamical} construction starts from a simple seed network on $S_g$ and then allows it to evolve by elementary moves that re-wire links and add or remove nodes under the constraint that the resulting network remains a simple graph embedded on $S_g$. 
This approach allows us to construct ensembles of graphs which share a common embedding. 
We find that the properties of the resulting graphs are related to the properties of $S_g$.
This is discussed in detail in the next section.

\begin{figure}
\vspace{-.5cm}
\begin{center}
\resizebox{0.85\textwidth}{!}{%
\includegraphics{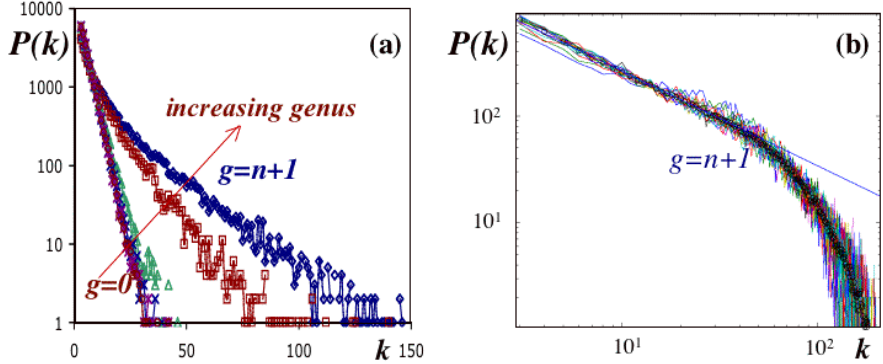}
}
\end{center}
\vspace{-0.4cm}
\caption{ 
\label{f.degrees}
{\bf a)} Degree distributions (log-normal scale) resulting from the numerical generation of triangulations on manifolds with different genera. 
{\bf b)} Degree distributions (log-log scale) for the case $g=n+1$ over several different simulations.
}
\end{figure}

\vspace{-0.8cm}
\section{Elementary Moves}\label{EM}
\vspace{-0.8cm}
Here we further restrict our investigations to triangular embeddings. Consider the effect of cascades of elementary moves of two types:  \emph{edge switching} (T1) and \emph{vertex insertion and removal} (T2) \cite{AsteSherr}.
 The first operation is a local elementary move which switches the connections among four nodes as indicated in Fig.\ref{f.T1T2}. The operation switches neighbors:
two first-neighboring nodes become second-neighbors ($i,j$ in the figure) whereas two second-neighbors become first-neighbours ($k,l$ in the figure).
This  $T1$ operation can be iterated allowing the exploration of a large class of triangular embeddings on $S_g$.
However, certain triangulations of oriented surfaces cannot be transformed into each other by  $T1$ operations alone. 
The second elementary operation ($T2$) is the insertion of a vertex in the middle of an existing triangle and its inverse, as drawn in Fig.\ref{f.T1T2}.
It is known that any triangulation on $S_g$ can be transformed into any other by a sequence of $T1$ and $T2$ elementary moves \cite{Alexander,Pach91}.
Applications of this technique to the case $g=1$ are discussed in \cite{AsteSherr,AsteHol,AsteBenoit}.

\vspace{-0.5cm}
\section{T1 networks}
\vspace{-0.5cm}
Consider next examples of large triangulations (containing up to $n = 10^5$ nodes) embedded on surfaces of various genera. 
For convenience, we consider values of $g$ between 0 and $n$+1, as the latter bound allows networks with integer connectivity (18) irrespectively of their size.
These networks are generated via the ``dynamical method'' introduced above. 
We start with a seed-graph and then apply $T1$ moves at random.
The result is a disordered network.
Remarkably, for the case $g=0$ the degree distribution is known analytically ($n \gg 1$) \cite{Brezin78,Boulatov86,Godrech92,Bouttier03}:  $P(k) =  16(3/16)^k (k-2) (2k-2)!/[k!(k-1)!] $, which, in the tail region, is well described by an exponential law $P(k) \sim P_0 exp(-\alpha k)$ with $\alpha \sim 0.3$.
Such theoretical behavior is confirmed numerically by our simulations using the `dynamical' construction.
At low genus, the degree distribution decreases exponentially with $k$ (linear trend in log-normal scale), as shown in Fig.\ref{f.degrees}(a).
However, it is also evident from the figure that when $g$ increases the coefficient $\alpha$ decreases and the distributions' tails become `fatter' .
Moreover, log-log plots of the degree distributions of networks embedded on high-genus surfaces reveal power-law behaviors for small-medium values of the connectivity (Fig.\ref{f.degrees}(b)).
We find in those cases very good fits of the distributions to the functional form: $P(k) \sim P_0 k^{-\beta} \exp(-\alpha k^{\gamma})$. (For the case in Fig.\ref{f.degrees} with $n=10000$ and $g=n+1$,  $\alpha \simeq 4 \; 10^{-4}$, $\beta \simeq 1$ and $\gamma \simeq 1.9$.)
This `stretching' of the degree distributions away from exponential behavior (characteristic of low genus) towards power-law behavior (with an exponential cutoff) suggests the presence of a phase-transition when the number of handles in the surface becomes comparable with the number of nodes.
Indeed, it was shown in \cite{AsteSherr} that the behavior at $g=1$ can be retrieved easily by modelling the system as non-interacting nodes constrained by a global condition on the average connectivity (which must be equal to 6 at $g=1$).
 On the other hand, networks embedded on topologically complex surfaces allow longer-range correlations to play a crucial role in the system organization, leading to non-Boltzmann distributions.

Another important effect associated with the variation of genus of the embedding surface concerns the intrinsic (or fractal) dimension of the network.
We infer this dimensional measure from the average length ($\left< j \right>$) of the geodesic paths on the network between any pair of nodes.
If this topological length  scales with the total number of nodes (the `volume') as $\left< j \right> \propto n^{1/d}$, one can define an intrinsic dimension $d$ for the system.
(This definition is analogous to standard scaling laws in $d$-dimensional lattices.)
Another possible definition of intrinsic dimensionality follows from analysis of the average number of nodes ($n_j$) at a given geodesic distance ($j$) from a central node \cite{AsBoRi}.
Indeed, in systems with finite intrinsic dimensionality, we expect  such a topological `perimeter' to grow with topological radius as: $n_j \propto j^{d-1}$.
Clearly, there are various possible definitions for the intrinsic dimension and, in general, they can lead to different results.
However we verified that the two definitions above lead to the same result for all the cases examined.
Interestingly, results associated with the application of the Regge calculus in 2D quantum gravity \cite{Ambjorn95} show that for $g=0$ the intrinsic dimension of random triangulations must be $d=4$.
We verified that our numerical simulations for the case $g=0$ are very well described with the functional forms  $\left< j \right> = c_1 n^{1/d} + c_2$ and  $n_j =  c_3 j^{d-1} + c_4$ with $d\simeq 4$. 
We have also verified that the same forms hold for all the small-genus cases (when $g \ll n$), again with $d \sim 4$.
Therefore we affirm that for small genus the system has a finite intrinsic dimension and it behaves as a \emph{large world}.
At the opposite limit we know that for very high genus -- when $g^*$ is neared and the network  approaches $K_n$ --  we attain 'saturation': $\left< j \right> \sim O(1)$. 
We have detected a critical region, in the vicinity of $g\sim n$, where the intrinsic dimension diverges.
In this region we observe  that the system becomes a \emph{small world} with: $n_j =  a_1 \exp(a_2 j) + a_3$  and  $\left< j \right> = a_4 \ln(n) + a_5$. 
We expect to observe another transition to ultra-small worlds ($\left< j \right> \sim \ln(\ln(n))$)  when the genus approaches $g^*$.
But this will be the topic of a forthcoming paper.

{\bf Acknowledgements}
T. Di Matteo benefited from discussion with the participants at the COST P10 ``Physics of Risk'' meeting in Nyborg (DK), April 2004. 
We acknowledge partial support from ARC Discovery Project DP0344004 (2003) and APAC.


\vspace{-0.8cm}
\begin{thebibliography}{}
\vspace{-0.8cm}


\bibitem {Amaral} 
L. A. N. Amaral, A. Scala, M. Barthelemy, and H. E. Stanley, 
{\it Proc. Natl. Acad. Sci.} {\bf{97}} (2000) 11149. 

\bibitem {Barabasi}
R$\acute{e}$ka Albert and Albert-L$\acute{a}$szl$\acute{o}$
Barab$\acute{a}$si, 
{\it Rev. Mod. Phys.} {\bf{74}}  (2002) 47. 

\bibitem {Newman} 
M. E. J. Newman, 
{\t SIAM Rev.} {\bf{45}} (2003) 167.

\bibitem{Cohen03}
R. Cohen and S. Havlin,
{\it Phys. Rev. Lett.} {\bf 90} (2003) 058701.

\bibitem{WattSt98}
D. J. Watts and  S. H. Strogatz, 
{\it Nature} {\bf 393} (1998) 440-442.

\bibitem{ER60}
 P.  Erdos and A. R\'enyi,  
{\it Publ. Math. Inst. Hungar. Acad. Sci.} {\bf 5} (1960)  17-61.

\bibitem{p16}	
H. M. Ohlenbusch, T. Aste, B. Dubertret and N. Rivier, 
{\it Eur. Phys. J. B}  {\bf 2} (1998) 211-220.

\bibitem{p17}	
B. Dubertret, T. Aste, H. M. Ohlenbusch and N. Rivier,  
{\it Phys. Rev. E} {\bf 58} (1998) 6368-6378.

\bibitem{AsteSherr}	
T. Aste and D. Sherrington, 
{\it J. Phys. A} {\bf 32} (1999) 7049-56. 

\bibitem{Ringel}
G. Ringel, 
``Map Color Theorem'', 
(Springer-Verlag, Berlin, 1974) cap. 4.

\bibitem{p14}
P. J. Gilbin, 
``Graphs, Surfaces and Homology'', 
(Chapman and Hall, 2nd edition, 1981).

\bibitem{Ringel1968}
G. Ringel and J. W. T. Youngs, 
{\it Proc. Nat. Acad. Sci. USA} {\bf 60} (1968) 438-445.

\bibitem{DM04interest}
T. Di Matteo, T. Aste, S. T. Hyde and S. Ramsden,
{\it preprint} (2004).

\bibitem{Michele}
M. Tumminello, T. Aste, T. Di Matteo, R. N. Mantegna, 
2004, in preparation.

\bibitem{p15} 	
J. Stillwell, 
 ``Classical Topology and Combinatorial Group Theory''‚
 (Springer-Verlag, 2nd edition, 1993). 

\bibitem{Cox}
H. S. M. Coxeter, 
``Regular Polytopes'' (Dover, New York, 1973).

\bibitem{HydeEUPJB03}
S. T. Hyde and S. Ramsden, 
{\it Eur. Phys. J. B}  {\bf 3178} (2003) 273-284.

\bibitem{HydeSSS03}
S. T. Hyde, S. Ramsden, T. Di Matteo, J. J. Longdell,
{\it Solid State Sciences} {\bf 5} (2003) 35-45.

\bibitem{Bonn00}
C. P. Bonnington, M. J. Grannell, T. S. Griggs, J. Siran,  
{\it Journal of Combinatorial Theory Series B} {\bf 78} (2000) 169-184.

\bibitem{Alexander}
 J. W. Alexander, 
{\it  Ann. of Math.} {\bf 31} (1930)  294--322.

\bibitem{Pach91}
 U. Pachner, 
{\it Arch. Math} {\bf  30} (1978) 89-91; 
U.  Pachner, 
{\it Europ. J. Combinatorics} {\bf 12} (1991) 129-145

\bibitem{New26} MHA Newman, 
{\it Royal Acad. Amsterdam} {\bf 29} (1926) 610-641.

\bibitem{AsteHol}
H. M. Ohlenbusch, T. Aste, B. Dubertret, and N. Rivier,
{\it Eur. Phys. J. B}  {\bf 2}  (1998) 211-220.

\bibitem{AsteBenoit}
B. Dubertret, T. Aste, H. M. Ohlenbusch and N. Rivier,
{\it Phys. Rev E} {\bf  58} (1996)  6368-78.

\bibitem{Brezin78}
E. Br\'ezin, C. Itzykson, G. Parisi, J. B. Zuber, 
{\it Comm. Math. Phys.}  {\bf 59}  (1978)  35.

\bibitem{Boulatov86}
D. V. Boulatov, V. A. Kazakov, I. K. Kostov, A. A. Migdal,
{\it Nucl. Phys. B} {\bf 275} (1986) 641.

\bibitem{Godrech92}
C. Godr\`eche, I. Kostov, I. Yekutieli, 
{\it Phys. Rev. Lett.} {\bf 69} (1992)  2674.

\bibitem{Bouttier03}
J. Bouttier, P. Di Francesco, E. Guitter,
{\it Nuclear Physics B} {\bf 675} (2003) 631.


\bibitem{AsBoRi}
T. Aste, D. Boos\'e, and N. Rivier,
{\it Phys. Rev E} {\bf  53} (1996)  6181-91.


\bibitem{Ambjorn95}
J. Ambjorn, Y. Watabiki,
{\it Nuclear Physics B} {\bf 445} (1995) 129-142;
J. Ambjorn, J. Jurkiewicz, Y  Watabiki,
{\it Nuclear Physics B} {\bf 454} (1995) 313-342.



\end{thebibliography}
\end{document}